\documentclass[sigconf,screen,nonacm]{acmart}

\AtBeginDocument{%
  \providecommand\BibTeX{{%
    \normalfont B\kern-0.5em{\scshape i\kern-0.25em b}\kern-0.8em\TeX}}}

\setcopyright{none}
\copyrightyear{2022}
\acmYear{2022}
\acmDOI{XXXXXXX.XXXXXXX}

\usepackage{subfigure}
\usepackage{multirow}
\usepackage{xcolor}

\usepackage{graphicx}
\graphicspath{ {./figs/} }
\usepackage{xspace}
\usepackage{tabularx}

\newcommand{\unisgga}{UniSGˆGA\xspace}
\newcommand{\rulesep}{\unskip\ \vrule\ }

\title[\unisgga: A GA-empowered Universal Scenegraph]{ \unisgga: A 3D scenegraph powered by Geometric Algebra unifying geometry, behavior and GNNs towards generative AI }

\begin{document}

\author{Manos Kamarianakis}
\email{kamarianakis@uoc.gr}
\orcid{0000-0001-6577-0354}
\affiliation{%
  \institution{FORTH - ICS, University of Crete, ORamaVR}
  \city{Heraklion}
  \country{Greece}
}

\author{Antonis Protopsaltis}
\orcid{0000-0002-5670-1151}
\email{aprotopsaltis@uowm.gr}
\affiliation{%
  \institution{University of Western Macedonia, ORamaVR}
  \city{Kozani}
  \country{Greece}
}

\author{Dimitris Angelis}
\orcid{0000-0003-2751-7790}
\email{dimitris.aggelis@oramavr.com}
\affiliation{%
  \institution{FORTH - ICS, University of Crete, ORamaVR}
  \city{Heraklion}
  \country{Greece}
}

\author{Paul Zikas}
\orcid{0000-0003-2422-1169}
\email{paul@oramavr.com}
\affiliation{%
  \institution{University of Geneva, ORamaVR}
  \city{Geneva}
  \country{Switzerland}
}

\author{Mike Kentros}
\orcid{0000-0002-3461-1657}
\email{mike@oramavr.com}
\affiliation{%
  \institution{FORTH - ICS, University of Crete, ORamaVR}
  \city{Heraklion}
  \country{Greece}
}

\author{George Papagiannakis}
\orcid{0000-0002-2977-9850}
\email{papagian@ics.forth.gr}
\affiliation{%
  \institution{FORTH - ICS, University of Crete, ORamaVR}
  \city{Heraklion}
  \country{Greece}
}

\renewcommand{\shortauthors}{Kamarianakis, Protopsaltis, Angelis et al.}

\keywords{Geometric Algebra, Generative AI, Graph neural networks, 3D scenegraph, Entity Component System, Behavior embedding}

\begin{abstract}
This work presents the introduction of \unisgga, a novel 
integrated scenegraph structure, that to
incorporates behavior and geometry data on a 3D scene. It is specifically designed to 
seamlessly integrate Graph Neural Networks (GNNs) and address the challenges associated with transforming a 3D scenegraph (3D-SG) during generative tasks. To effectively capture and preserve the topological relationships between objects in a simplified way, within the graph 
representation, we propose \unisgga, that seamlessly integrates Geometric Algebra (GA) forms.
This novel approach enhances the overall performance and capability 
of GNNs in handling generative and predictive tasks, opening up new possibilities and aiming to lay the foundation for further exploration and development of 
graph-based generative AI models that can effectively 
incorporate behavior data for enhanced scene generation and 
synthesis.

\end{abstract}

\maketitle

\section{Introduction}

The recent success of pre-trained foundation models, such as GPT (Generative Pre-trained Transformer), has paved the way for evolution in geometric deep learning \cite{bronstein2017geometric} and GNNs \cite{hu2019strategies}. Such advancements have greatly improved the generation of static 3D scenes \cite{dhamo2021graph} by incorporating relational patterns within the graph topology as node or link features. Typically, these scenes rely on well-defined 3D-SGs. The creation of immersive VR experiences require the incorporation of behavioral information and interactions, that are specified with the adoption of the graph structure Lessons-Stages-Actions (LSA) \cite{zikas2023mages4}.

\begin{figure*}[tb]
    \centering
    \includegraphics[width=\textwidth]{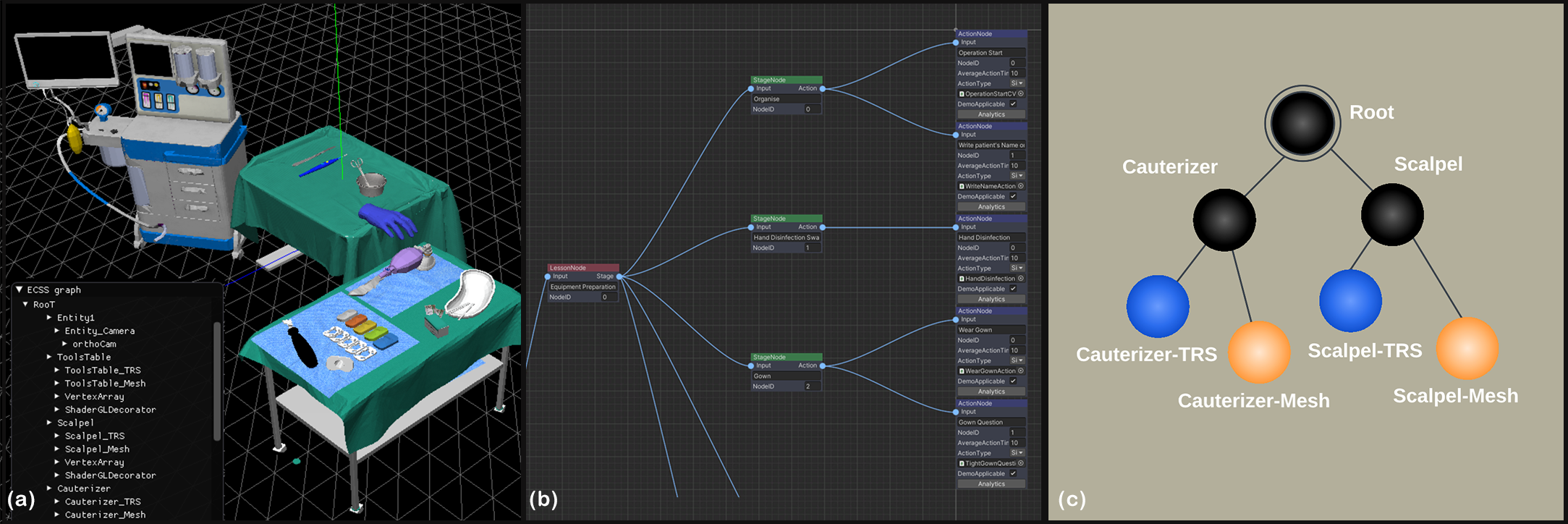}
    \caption{The 
    \unisgga unifies the three diverse 
    graphs that must be maintained for a 3D scene
    that includes behaviour, digestible 
    by a 
    GNN: (a) the 3D scenegraph with Entity-Component-Systems, (b) the behavioral 
    LSA graph and (c) the deriving GNN graph. The components 
    describing the parent-child relative topology are expressed
    in GA-based forms, for increased performance on 
    predictive and/or generative tasks.}
    \Description{ }
    \label{fig:three_graphs}
\end{figure*}

Nevertheless, the efficient input of all encapsulated data to GNNs poses a challenge, as it requires managing three distinct graph structures (see Figure~\ref{fig:three_graphs}), namely 3D geometry, interactive event-based animations encapsulated as behaviours (LSAs) and GNNs. This introduces a great complexity in maintaining transformations between these graphs that may lead to a potential bottleneck. To address these limitations, we propose the Universal Scenegraph (UniSG), a novel data structure aimed at providing a no-code approach featuring GNNs, that generate new nodes, edges, and features, reflecting the creation of 3D models, scenes, and behavioral steps. UniSG paves the way towards generative AI techniques, by integrating Entities-Components-Systems (ECS), 3D-SGs, and LSAs with GNNs, simplifying the creation of 3D scenes with embedded behavior, and mitigating existing process bottlenecks.

UniSG leverages a representation form that is able to capture and preserve relative topological information between parent and child entities. Rather than relying on conventional Euclidean-based matrix form or Euler angles or dual/single quaternions, commonly employed in 3D scenes, we utilize Geometric Algebra (GA) based forms, such as multivectors; the resulting model is denoted as \unisgga. GA-based representations 
enable the encapsulation of diverse transformation 
data in a unified format, facilitating 
\cite{Pepe.2023}
deeper geometric connections, thereby influencing the performance of GNNs across various tasks
(see Section~\ref{sub:GA_GNNs}).

\subsection{The importance of GNNs for Generative AI}
\label{sub:GNNs}

GNNs have gained significant attention in recent years due to their effectiveness in handling graphs of varying types, sizes, structures, connectivity patterns and data with complex relational structures, due to the high flexibility and adaptability of their architectures. Their design makes them particularly well suited for generative and predictive AI tasks that involve graph-structured data, like complex 3D scenes, where nodes represent objects and edges encode relationships or connections between them, as they are able to capture spatial relationships, model dependencies and extract meaningful representations. Specifically for an entity-component-systems (ECS) in a scenegraph CG framework \cite{papagiannakis2023elements,UniSG2023}, a GNN involves heterogeneous nodes, representing entities and diverse components, containing object-related data (transform, mesh, image texture data, etc.).

GNN aggregation allows the capture of the graph's local dependencies, while its propagation through the graph allows the capture of global dependencies. In this context, complex interactions between nodes may also be captured by iterative node representation refinement, using message-passing mechanisms.  Such rich information about the nodes and their spatial relationships, learned from the training data, may be encoded in meaningful and low-dimensional embeddings, that involve fixed-length vectors or a continuous feature space. The GNN model may be trained in a) supervised manner, involving annotated 3D-SGs, aiming to predict missing elements or labels, and b) unsupervised manner involving graph similarity or reconstruction losses, aiming to optimize the generative model.

\subsection{GA and GNNs}
\label{sub:GA_GNNs}

The combination of GA with 
GNNs offers several benefits across different domains 
and tasks\cite{GeometricNeuralComputingbayro2001}.
GA-based approaches have demonstrated superior 
information (inherent structures and correlations among multiple dimensions) preservation, as multi-dimensional data are represented through multivectors. This leads to improved 
performance, compared to traditional techniques, in tasks including as 
time series processing, hyperspectral image analysis, and traffic 
prediction \cite{Zang.20227nf,Miao.2022,Oktar.2022,LearningShapeMotionRepresentationsli2019,Liu.2022}. They also exhibit reduced overfitting risks, compared to real-valued counterparts, making them more effective in capturing complex features while maintaining the 
multi-dimensionality of the data.

GA is particularly advantageous in handling rotational data, 
making it valuable for computer vision tasks, like pose estimation or protein prediction \cite{Pepe.20225er,Pepe.2023}. 
GA-based formulations enable better regression on rotations and 
can reduce errors in high-noise datasets while learning fewer 
parameters. Additionally, GA-based graph feature embedding 
enhances the quality and presentation of graph features in GNNs. By leveraging the high algebraic dimensions 
of GA, feature information distortion across hidden layers can be 
minimized, resulting in improved performance in graph-related tasks.
Furthermore, GA-based approaches can 
reduce computational complexity by utilizing appropriate multivector 
representations and exploiting the algebraic properties of GA. 
This reduction in complexity enables more efficient data processing 
and analysis, with fewer parameters to be learned without 
compromising performance.

In summary, the integration of GA with Neural 
Networks offers benefits, such as enhanced representation of 
multi-dimensional data, improved information preservation, 
effective handling of rotational data, better graph feature 
embedding, robustness to poor network conditions, and reduction 
of computational complexity. These advantages make GA a 
valuable framework for various scientific domains and tasks, 
facilitating more accurate and efficient data processing and 
analysis.

\par{\textbf{Paper Overview.}} In Section~\ref{sec:unisg} we introduce 
the UniSG model, whereas in Section~\ref{sec:unisg_ga} we 
propose the enhanced \unisgga model that exploits GA-based 
representation forms. These models are implemented and available 
to use within the Elements project, which now includes 
enhanced GA-functionalities, as described in Section~\ref{sec:elements}. 
Results obtained for our models performance are presented in 
Section~\ref{sec:results}, followed by Conclusions, Future Work 
and Acknowledgments.


\section{UniSG: A Universal SceneGraph} 
\label{sec:unisg}

The UniSG system, introduced in a concise manner in 
\cite{UniSG2023}, exhibits a heterogeneous graph structure 
built upon the Entity Component System in a Scenegraph 
(ECSS) model, such as the one proposed in \cite{papagiannakis2023elements}. 
This graph encompasses diverse component types capable 
of storing both geometric and behavioral information 
relevant to interaction with the 3D scene and events 
triggered by specific conditions. Specifically, the UniSG 
graph incorporates three types of components: \texttt{info}, \texttt{TRS}, 
and \texttt{mesh}. The \texttt{info} components maintain a count of node 
types among their children, while the \texttt{TRS} components store 
a 16-dimensional vector obtained by flattening the corresponding 
transformation matrix. 
The \texttt{mesh} components house a feature vector of size 1024, 
representing the mesh using a suitable encoder such as 
the AtlasNetEncoder \cite{Groueix_2018_CVPR} combined with 
a Poisson sampling process. This encoding methodology ensures 
a fixed-size representation regardless of the complexity of 
the original mesh. Subsequently, the resulting vector can be 
decoded using the AtlasNetDecoder to generate a point cloud, 
which can then be further reconstructed into a triangulated mesh.

To incorporate behavioral functionality, the UniSG system 
introduces a forth \texttt{ActionData} component that stores data 
pertaining to desired behavioral characteristics, accompanied 
by appropriate \texttt{Action} systems responsible for processing 
this data. These ECS components and systems effectively 
represent user actions required within a training scenario, 
akin to those stored in the Lesson-Stages-Actions (LSA) 
data structure \cite{zikas2020immersive}. 
The \texttt{ActionData} nodes adhere to a standardized structure for 
all actions and store action-specific data and conditions in 
vector form. The diverse \texttt{Action} systems continuously traverse 
the graph or its designated sections to validate whether the 
specified conditions are met.

\begin{figure*}[tb]
    \centering
    \includegraphics[width=0.62\textwidth]{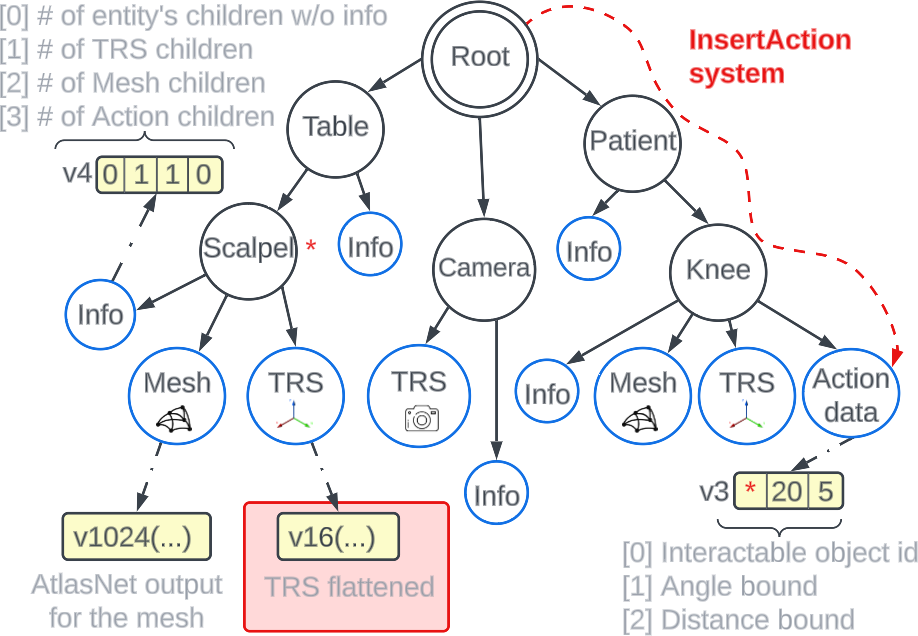}
    \rulesep
    \includegraphics[width=0.33\textwidth]{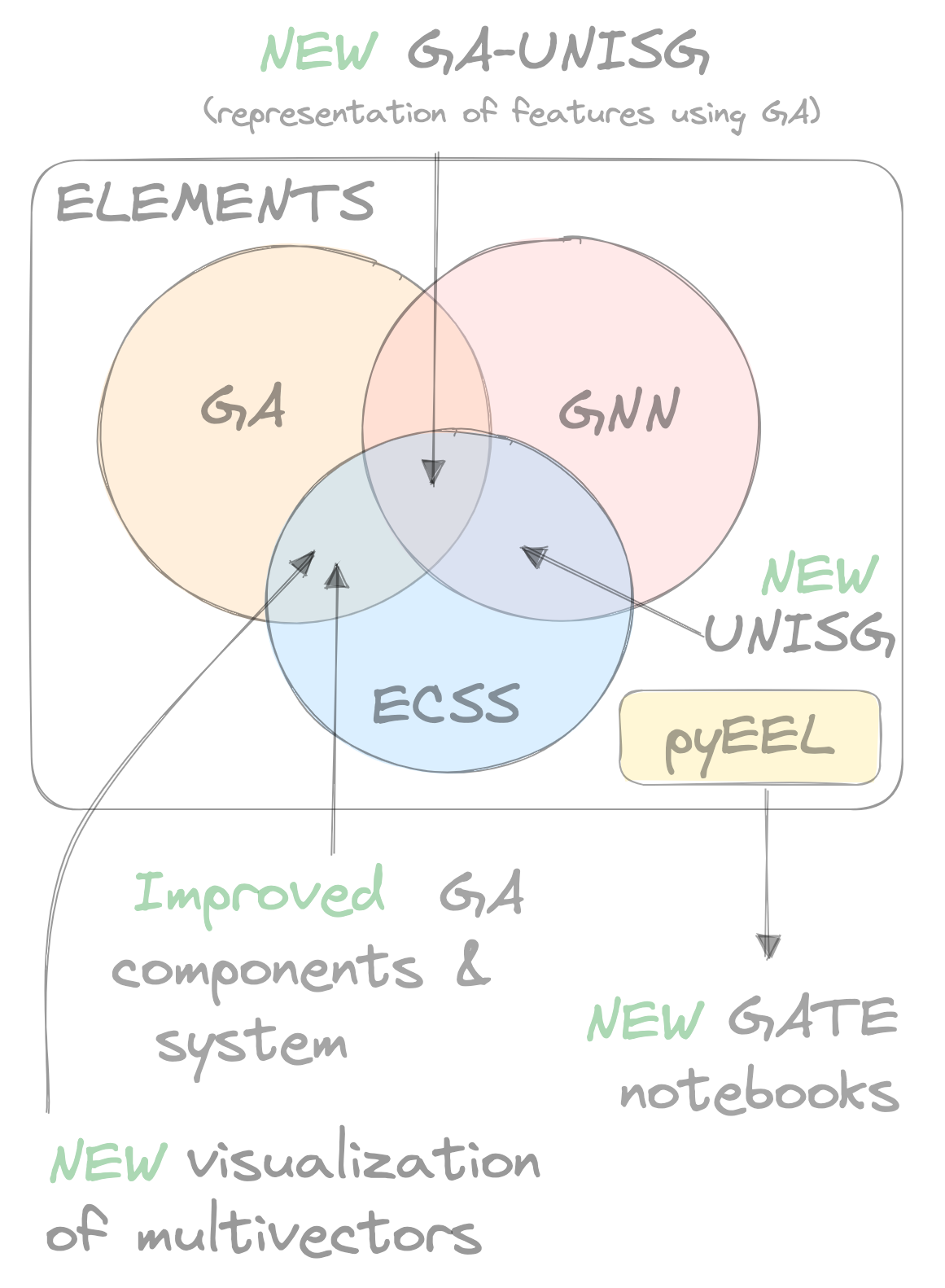}
    \caption{(Left) As opposed to the
    UniSG model \cite{UniSG2023}, the proposed \unisgga model 
    suggests using any GA-based representation form for the \texttt{TRS} component (red box), instead of the original 16-dimensional 
    array vector, deriving from the flattening of a transformation 
    matrix. (Right) A diagram denoting the contributions 
    presented in this paper, with respect to state-of-the-art. }
    \Description{ }
    \label{fig:unisgga}
\end{figure*}

The architectural elements of the ECS framework are depicted in 
Figure ~\ref{fig:unisgga} as follows. The black nodes represent 
entities, while 
the blue nodes represent components, which encapsulate various 
data such as transformations, meshes, and actions. 
Systems, represented by red lines, process the data contained 
in components and perform specific tasks while traversing the 
graph. Graph features, highlighted in yellow, are represented in vector form, enabling their utilization by GNNs for further analysis and processing.

Figure ~\ref{fig:unisgga} also exemplifies the implementation 
of an "Insert" action within the UniSG  system. In this specific 
scenario, the \texttt{InsertAction} 
system is responsible for verifying whether the placement 
of the scalpel on the knee adheres to the specified spatial 
boundaries. This check is performed when the system visits 
the \texttt{ActionData} component.

To consolidate disparate data types into a unified format, 
various file formats commonly employed have been merged into 
a single master file. Pixar's Universal Scene Description 
(USD) (\url{http://graphics.pixar.com/usd/}) future-proof format has been selected for 
its exceptional versatility, enabling the inclusion of more 
advanced features such as VR-Recording \cite{VRRR}.

\section{\unisgga: Empowering UniSG with Geometric Algebra} 
\label{sec:unisg_ga}

The original UniSG model employed a \texttt{TRS} component, which stored 
the topological relationship between an entity and its parent as 
a 16-dimensional array vector. 
This vector was obtained by flattening a 4x4 transformation 
matrix, resulting from the multiplication of Translation, Rotation, 
and Scaling matrices.

In this paper, we propose the \unisgga model, which overcomes 
the limitation of relying solely on matrix-derived vectors. 
The \unisgga model suggests the utilization of alternative 
forms of transformation data, allowing for a more diverse 
range of representations. Particularly, we advocate for the 
adoption of GA to express data that 
represents geometrical relationships. The integration of GA is 
not merely intended to promote its acceptance, but rather to 
demonstrate its potential to yield improved outcomes in various 
scientific domains, particularly those involving predictive and 
generative tasks, with a special focus on GNNs.


\section{\unisgga within the Elements project} 
\label{sec:elements}

The proposed \unisgga structure is already implemented within
the Elements project, introduced in \cite{papagiannakis2023elements}, similar 
to its predecessor UniSG.
Elements, presents 
a pioneering open-source pythonic framework based on 
entity-component-systems (ECS) implemented within a scenegraph architecture. 
It is explicitly tailored to address the demands of scientific, visual, and 
neural computing applications. Comprised of three vital Python components—pyECSS, 
pyGLV, and pyEEL—the Elements package offers a foundational implementation of the 
ECS paradigm, accompanied by practical examples that proficiently familiarize 
even inexperienced computer graphics programmers with fundamental principles 
and methodologies. Notwithstanding its straightforwardness, Elements retains a 
transparent nature, affording users the ability to scrutinize and manipulate 
each stage of the graphics pipeline. Leveraging Python's inherent advantages 
in rapid prototyping and development, users can augment Elements' capabilities 
by introducing novel components and systems or refining existing ones. 

The collection of jupyter notebooks within the pyEEL repository serves as a 
demonstrative repository for showcasing the influence of Elements' present and 
future features across diverse scientific domains and packages, thereby establishing 
a valuable pedagogical resource for both novice and intermediate developers.
To facilitate the transition to GA forms, pyEEL now
incorporates a series of Jupyter notebooks that serve three 
purposes: (a) introducing basic 
GA concepts to users unfamiliar with GA, (b) demonstrating 
the equivalence between different representation forms in a 
digestible manner for intermediate GA users, and (c) presenting 
more advanced applications of these principles, such as model 
animation using GA, for experienced GA users.

\subsection{Geometric Algebra powered 3D scenegraph}

Currently, matrix representations dominate the field due to 
their ease of implementation and compatibility with GPU 
shader-level operations. Although quaternions have mitigated 
issues such as gimbal lock and interpolation artifacts when 
evaluating rotation matrices, GA introduces a further advancement 
in representation forms. By utilizing translators, rotors, and 
dilators as GA-based counterparts for translation, rotation, 
and dilation, respectively, we can achieve improved results 
both quantitatively (reducing the number of keyframes required for 
interpolation) and visually \cite{LessIsMore}.

Complex operations, such as 
extracting geometric information from motors (i.e., geometric 
products of a translator 
and a rotor), are now performed with ease, by
leveraging the capabilities of the 
well-maintained Clifford Python package 
\cite{clif}, facilitating efficient transmutation between different forms.

Specifically, let $M$ be a 4x4 matrix representing a rotation
followed by translation. It is well known that the top 
left 3x3 submatrix is a rotation matrix  and the 3 first elements
of the last column is the translation vector $t$. From $R$
matrix we can  extract the angle/axis, and therefore determine
the equivalent unit quaternion $q$ that expresses the same rotation. 
Finally, having the quaternion and the translation vector you 
can easily concatenate them to obtain the respective 
dual-quaternion $dq$. The following is summarized in 
\eqref{eq:equivalent_euclidean}, where rotational data 
are represented in cyan, translational in blue and mixed data in purple.
\begin{align}\label{eq:equivalent_euclidean}
\begin{split}
\textcolor{purple}{M} = 
\begin{bmatrix}
\textcolor{cyan}{m_{1}} & \textcolor{cyan}{m_{2}} & \textcolor{cyan}{m_{3}} & \textcolor{blue}{t_1} \\
\textcolor{cyan}{m_{4}} & \textcolor{cyan}{m_{5}} & \textcolor{cyan}{m_{6}} & \textcolor{blue}{t_2} \\
\textcolor{cyan}{m_{7}} & \textcolor{cyan}{m_{8}} & \textcolor{cyan}{m_{9}} & \textcolor{blue}{t_3} \\
0 & 0 & 0 & 1
\end{bmatrix}
\Leftrightarrow
R = \begin{bmatrix}
\textcolor{cyan}{m_{1}} & \textcolor{cyan}{m_{2}} & \textcolor{cyan}{m_{3}} \\
\textcolor{cyan}{m_{4}} & \textcolor{cyan}{m_{5}} & \textcolor{cyan}{m_{6}} \\
\textcolor{cyan}{m_{7}} & \textcolor{cyan}{m_{8}} & \textcolor{cyan}{m_{9}} 
\end{bmatrix} \& \ 
\textcolor{blue}{t = (t_1, t_2, t_3)} 
\\
\Leftrightarrow 
\textcolor{cyan}{(\text{Angle},\text{Axis})}\  \& \  \textcolor{blue}{t} 
\Leftrightarrow 
\textcolor{cyan}{\text{Quaternion } q}\ \& \ \textcolor{blue}{t}
\Leftrightarrow
\textcolor{purple}{\text{Dual-Quaternion}}\  \textcolor{purple}{dq}.
\end{split}
\end{align}

From the translation vector $t$, we can easily determine the 
corresponding translator $T_{PGA}$ in 3D PGA as follows:
\begin{align}
T_{PGA} = 1 -0.5e'_0(t_1e'_1+t_2e'_2+t_3e'_3),
\end{align}
where $e'_0,e'_1,e'_2$ and $e'_3$ are basis vectors of 3D PGA.
Similarly, we can derive the 
corresponding translator $T_{CGA}$ in 3D CGA as :
\begin{align}
T_{CGA} = 1 -0.5e_0(t_1e_1+t_2e_2+t_3e_3)(e_4+e_5),
\end{align}
where $e_0,e_1,e_2, e_3,e_4$ and $e_5$ are basis vectors of 3D CGA. 
Extraction of the vector $t$ from both $T_{PGA}$ and $T_{PGA}$ is apparent as long as the multivectors are normalized; otherwise, 
a division by the scalar part is initially required. 

Given a unit quaternion $q=q_0 + q_1\pmb{i}+q_2\pmb{j}+q_3\pmb{k}$, we can easily determine the 
respective rotor $R_{PGA}$ in 3D PGA and
$R_{CGA}$ in 3D CGA (see \cite[Section~2.4]{LessIsMore}) as
\begin{gather}
R_{PGA} = q_0 - q_3e'_{12} +q_2e'_{13} - q_1e'_{23}, \text{ and } \\
R_{CGA} = q_0 - q_3e_{12} +q_2e_{13} - q_1e_{23},
\end{gather}
where $\{e'_{12}, e'_{13}, e'_{23}\}$ and 
$\{e_{12}, e_{13}, e_{23}\}$ are respectively PGA and CGA 
basis vectors. In conclusion,  the following 
equivalencies holds:
\begin{align}
\textcolor{blue}{T_{PGA}} \Leftrightarrow \textcolor{blue}{t} 
\Leftrightarrow \textcolor{blue}{T_{CGA}}, \text{ and \ \ }
\textcolor{cyan}{R_{PGA}} \Leftrightarrow \textcolor{cyan}{q} 
\Leftrightarrow \textcolor{cyan}{R_{CGA}}.
\end{align}
Lastly, in \cite[Section~2.4]{LessIsMore}, it is shown that
given a PGA motor $M_{PGA}$ resulting from the geometric product 
of $T_{PGA}$ and $R_{PGA}$, one may extract the latter two.  
The same holds for a CGA motor $M_{CGA}$ resulting from the geometric product of the translator $T_{CGA}$ and the
rotor $R_{CGA}$, 
yielding:
\begin{align}
\textcolor{purple}{M_{PGA}} \Leftrightarrow \textcolor{cyan}{R_{PGA}}\ \&\ \textcolor{blue}{T_{PGA}}, , \text{ and \ \ }
\textcolor{purple}{M_{CGA}} \Leftrightarrow \textcolor{cyan}{R_{CGA}}\ \&\ \textcolor{blue}{T_{CGA}}.
\end{align}
Using all equivalencies described above we can now extend 
\eqref{eq:equivalent_euclidean} to the complete equivalency list of representation forms; all equivalencies can occur using 
functions implemented within the Elements framework:
\begin{gather}\label{eq:equivalent_complete}
\textcolor{purple}{\text{Transformation Matrix }M} 
\Leftrightarrow
\textcolor{cyan}{\text{Rotation Matrix } R }\  \& \ 
\textcolor{blue}{\text{vector }t} 
\Leftrightarrow \nonumber
\\
\textcolor{cyan}{(\text{Angle},\text{Axis})}\  \& \  \textcolor{blue}{t} 
\Leftrightarrow 
\textcolor{cyan}{\text{Quaternion } q}\ \& \ \textcolor{blue}{t}
\Leftrightarrow
\textcolor{purple}{\text{Dual-Quaternion}}\  \textcolor{purple}{dq}. \nonumber
\\
\Leftrightarrow 
\textcolor{purple}{M_{PGA}} \Leftrightarrow \textcolor{cyan}{R_{PGA}}\ \&\ \textcolor{blue}{T_{PGA}} 
\Leftrightarrow 
\textcolor{purple}{M_{CGA}} \Leftrightarrow \textcolor{cyan}{R_{CGA}}\ \&\ \textcolor{blue}{T_{CGA}}. 
\end{gather}

\begin{figure*}[tb]
   \centering
   \includegraphics[width=0.49\textwidth,height=145pt]{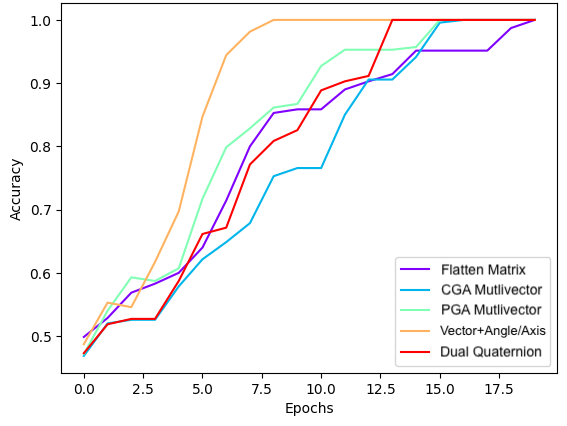}
   \includegraphics[width=0.49\textwidth,height=145pt]{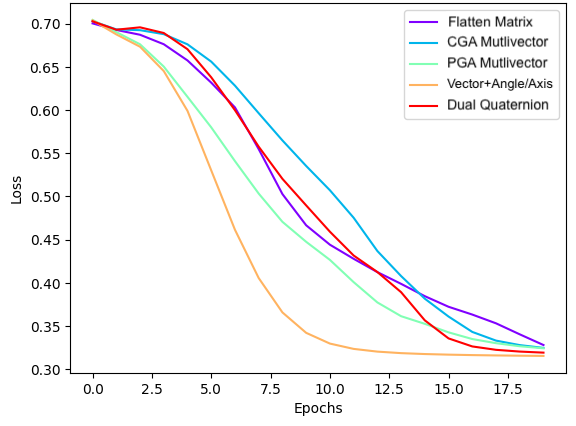}
   \caption{Train accuracy (Left) and Loss (Right), for the classification task described in Section~\ref{sub:classification}. These are mean values after running the experiment 10 times.}
   \Description{ }
   \label{fig:Classification}
\end{figure*}


\section{Results} 
\label{sec:results}

To validate the effectiveness of our proposed approach, 
we conducted three experimentation tasks in the domains of 
classification, generative modeling, and topology prediction for 
3D scenegraphs. In 
Figures~\ref{fig:Classification} and \ref{fig:generative_topology_prediction}, we present the obtained results 
using different representation forms for the \texttt{TRS} 
component of the \unisgga model. Specifically, we compare the use 
of a) flatten matrices (representing the original UniSG), b) CGA 
and c)PGA multivectors, d) a vector for translation combined with 
an angle and an axis for rotation , as well as a e) dual-
quaternion representation.

Each task is accompanied by a comparison graph, demonstrating 
the performance of the GA-based representations in relation to the 
conventional Euclidean-oriented formats. The results consistently 
show that the utilization of GA-based representation forms, such 
as CGA/PGA multivectors and dual-quaternions, either outperforms 
or performs on par with the traditional flatten matrices 
representation.

\subsection{Classification} 
\label{sub:classification}

Our methodology was evaluated through a classification task 
involving a neural network architecture composed of two 
Convolutional layers. The GraphSAGE convolution operation was 
applied to the input graph within this framework. To assess the 
performance of our approach, we curated a dataset comprising of 
100 3D scenes. These scenes were generated using a random noise-
based data augmentation technique, which involved perturbing the 
components of two behaviorally rich 3D scenes modeled using both 
the UniSG and \unisgga system. The scenes selected for 
augmentation were a surgical operating room (OR) and a living 
room. The dataset was split into training and testing sets, with a 
ratio of 70\% for training and 30\% for testing. The neural 
network model was trained for 20 epochs, and the GNN attention 
mechanism was employed. In the experimentation phase of our 
approach, we performed 10 runs for each experiment, which, 
remarkably, achieved a 100\% accuracy on both the training and 
testing splits, demonstrating its effectiveness.

In the experimentation results of the classification task, 
depicted in Fig~\ref{fig:Classification}, we notice a low initial 
mean accuracy on all methods, indicating a possible need for 
longer training or model adjustments. Accuracy improves 
consistently over epochs, exhibiting a few fluctuations in CGA and 
PGA. The steepness of the Vector+Angle/Axis curve indicates that 
the model learns quickly as its accuracy get 100\% after 7.5 
epochs. All curves seem to be converging to 100\% accuracy after 
17 epochs, a clear sign that it is performing well on the training 
data. 
We also notice a low initial loss on 
all curves, with vector+Angle/Axis curve to be minimizing faster, 
after 10 epochs, than the others. All loss curves seem to converge 
after 18 epochs, indicating a well performing model.


\subsection{Generative AI using UniSGˆGA} 
\label{sub:generative}

Our approach was further tested on a generative task. For 
this purpose, we generated a dataset of 1000 unique scenes with 
meaningful layouts, specifically representing a surgical operating 
room (OR). These scenes were then utilized to train a Conditional 
Graph Variational AutoEncoder (CGVAE). The primary objective of 
the CGVAE is to enable the addition of objects to an existing or 
empty scene based on their category, either sequentially or in 
bulk. Ultimately, since the utilized \unisgga structure includes behavior 
components, for all object entities, and the respective systems, we 
aim to train our autoencoder with scene objects that incorporate behavior and 
provide a complete generative AI solution (currently only topology generation is evaluated).

\begin{figure*}[tb]
    \centering
    \includegraphics[width=0.99\textwidth]{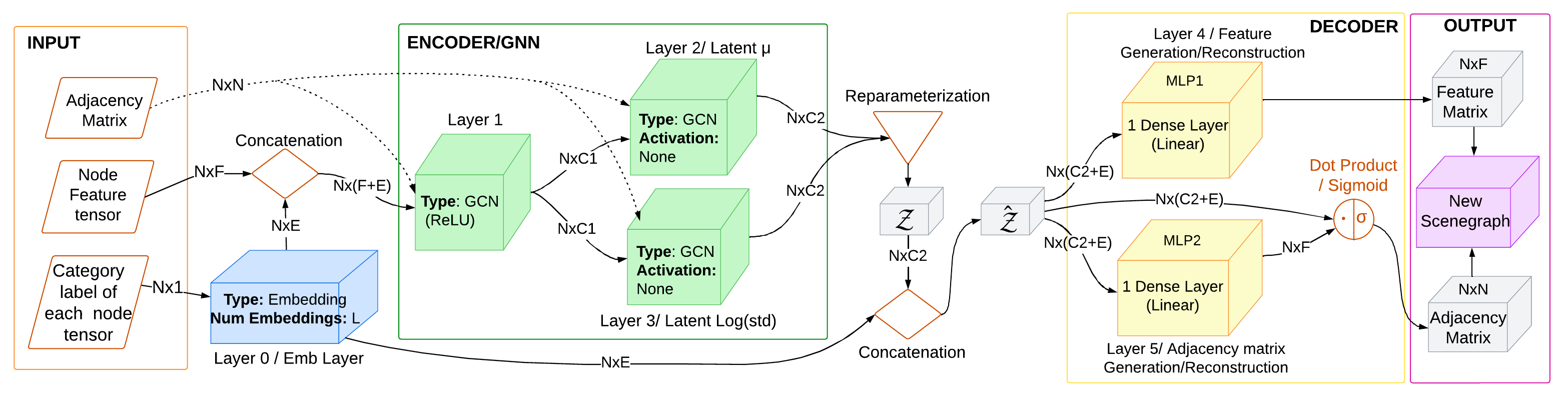}
    \caption{Diagram describing the generative task of Section~\ref{sub:generative}. }
    \Description{ }
    \label{fig:generative-diagram}
\end{figure*}

To achieve this, each entity node within the 
\unisgga was labeled with its corresponding category, e.g.,  
"Scalpel". During the training process, the Encoder module, which 
encompasses a GNN with Graph Convolutional 
layers, encodes the $N$ nodes of the graph using their inherent 
$F$ features and their associated category embeddings. For each of the 
nodes a vector $E$ is produced, by passing the labels through the embeddings, 
resulting in a $N$x$E$ matrix. The resulting encodings/latent space representation 
for each node, $\mathcal{Z}$, are concatenated with their respective category embeddings, by 
concatenating the input graph node matrix, of size $N$x$F$, with the embeddings, 
resulting in a $N$x($F$+$E$) matrix. This concatenated representation, denoted 
as $\mathcal{Z}$, is subsequently fed into the Decoder module, which consists of 
two Multilayer Perceptrons (MLPs): one for decoding 
the node features from $\hat{\mathcal{Z}}$ and one for decoding the 
adjacency matrix from $\hat{\mathcal{Z}}$ (see Figure~\ref{fig:generative-diagram}).

\begin{figure*}[tb]
   \centering
   \includegraphics[width=0.49\textwidth]{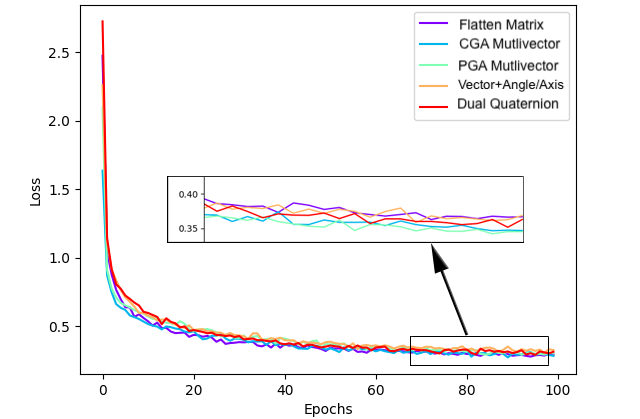}
   \includegraphics[width=0.49\textwidth]{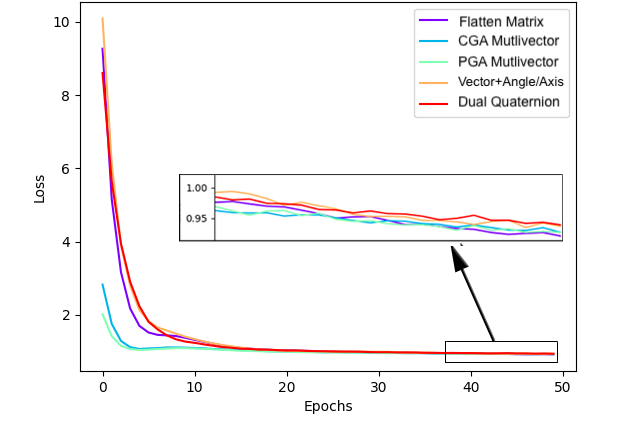}
   \caption{  
   Loss of epochs 0-100, regarding the generative task described in Section~\ref{sub:generative} (Left) and the topology (edge) prediction task described in Section~\ref{sub:topology_prediction} (Right).
   These are mean values after running the experiment 10 times.}
   \Description{ }
   \label{fig:generative_topology_prediction}
\end{figure*}

Our training procedure incorporates several loss 
functions. Specifically, we employ mean squared error (MSE) 
loss for node feature reconstruction, binary cross-entropy 
(BCE) loss for adjacency matrix reconstruction, and 
Kullback-Leibler (KL) divergence loss to encourage 
diversity in scene generation. 
As the model is conditionally trained using these categories, a conditional sampling of the generated scenes is possible, based on specific object categories. This allows the generation of scenes that are greatly influenced by the categories of existing or newly introduced nodes.

The experimentation results of the generative task, 
we see the loss in Fig~\ref{fig:generative_topology_prediction} (Left), 
that depicts the discrepancy between the generated 
and the target output. In this regard we notice 
that all mean losses are initially relatively low, 
with PGA and CGA significantly lower, meaning that 
all models produce high-quality outputs from the start. 
All loss reductions are minimized rapidly consistently 
below 1.0. Although all methods seem to converge very 
early, CGA and PGA mean loss curves are always below 
the others; which is indicative that the model is well-performing 
and that it has learned the generative task.


\subsection{Topology prediction} 
\label{sub:topology_prediction}

Finally, a topology prediction task was utilized to 
further evaluate the differences between UniSG and 
the GA-empowered \unisgga.
In such tasks, it is common to seek accurate predictions 
regarding the spatial relationships between objects, 
including relationships such as "above", "below", 
"right-of", as well as higher-level relationships like 
"part-of" or "connected-to". Our approach was 
specifically evaluated on a topology prediction task 
involving the identification of the "on-top-of" 
relationship between two objects. To address this 
prediction task, we made modifications to our previous 
model by transforming the Graph Variational AutoEncoder 
into a simplified Graph AutoEncoder that focused 
on adjacency matrix reconstruction for predicting 
the desired topology link based on the graph structure.
It is worth noting that while our modified model proves 
effective for certain topology prediction tasks, it may 
not capture the complexity of relationships or high-level 
semantics within the UniSG.

The experimentation results of the topology prediction 
task, depicted in Fig~\ref{fig:generative_topology_prediction} (Right), show 
that mean loss (on 10 runs) with CGA and PGA are initialy low and are 
minimized rapidly compared to other methods. Although 
all methods seem to converge early, after 15 epochs, CGA 
and PGA mean loss curves are always below the others, 
indicating a well-performing model. The loss curves do not 
show any signs of overfitting which is a direct consequence 
of the performed data augmentation, increasing diversity and 
quantity, of the training samples.
For each of the 10 runs, a single random scene was generated 
with 10000 cubes, and link prediction was performed on each run on a single scene.



\section{Conclusions and future work}

In this work, we introduced \unisgga, an integrated graph structure designed to be seamlessly compatible with Graph Neural Networks (GNNs) while incorporating behavior data. A key contribution of \unisgga is its ability to overcome the challenges associated with transforming a 3D scenegraph (3D-SG) when conducting generative tasks. By leveraging GA forms, \unisgga effectively captures and stores the topological relations between objects within the graph, while enhancing the performance and capability of GNNs when handling predictive and generative tasks.
This advancement paves the way for more efficient and intuitive approaches in generating complex 3D scenes with embedded behavior.

As a future endeavor, our plan is to train the GNN architecture 
of \unisgga using an extensive corpus of 3D scenes encompassing 
both content and behavior. This training dataset will consist of 
various types of scenes, including models and even segments of 
educational curricula. Through this training process, we aim to 
evaluate the performance of UniSG on intricate generative AI 
tasks, with the ultimate objective of enabling the generation of 
behavior-embedded 3D scenes in a streamlined manner, towards a no-code authoring pipeline.

\balance

\begin{acks}\label{sec:acks}

The project was partially funded by 
the National Recovery and Resilience Plan "Greece 2.0" - NextGenerationEU, under grant agreement No TA$\Sigma\Phi$P-06378 (REVIRES-Med), and Innovation project Swiss Accelerator under grant agreement 2155012933 (OMEN-E), supported by Innosuisse. 
\end{acks}

\bibliographystyle{ACM-Reference-Format}
\bibliography{references}

\end{document}